\documentclass[aps,prl,reprint,twocolumn,superscriptaddress,floatfix,nofootinbib,longbibliography]{revtex4-2}
\usepackage{graphicx,amsmath,amsfonts,amssymb,amsthm,xr}
\usepackage{epsfig,amsmath,amssymb,color,dsfont,upgreek,physics}
\usepackage{mathrsfs}
\usepackage{mathtools}
\usepackage{bbold}
\usepackage{float}
\usepackage[caption=false]{subfig}
\usepackage[bookmarks=true,colorlinks,linkcolor=OrangeRed,urlcolor=NavyBlue,citecolor=RoyalBlue]{hyperref}
\usepackage[capitalize]{cleveref}
\usepackage[dvipsnames]{xcolor}
\usepackage{orcidlink}

\definecolor{mygold}{rgb}{0.5,0.6,0.7}

\begin{document}

\title{Finite-Temperature Dynamical Phase Diagram of the $2+1$D Quantum Ising Model}

\author{Lucas Katschke}
\affiliation{Department of Physics and Arnold Sommerfeld Center for Theoretical Physics (ASC), Ludwig Maximilian University of Munich, 80333 Munich, Germany}
\affiliation{Munich Center for Quantum Science and Technology (MCQST), 80799 Munich, Germany}

\author{Roland C.~Farrell${}^{\orcidlink{0000-0001-7189-0424}}$}
\affiliation{Institute for Quantum Information and Matter, California Institute of Technology}
\affiliation{Department of Physics, California Institute of Technology}

\author{Umberto Borla${}^{\orcidlink{0000-0002-4224-5335}}$}
\affiliation{Max Planck Institute of Quantum Optics, 85748 Garching, Germany}
\affiliation{Department of Physics and Arnold Sommerfeld Center for Theoretical Physics (ASC), Ludwig Maximilian University of Munich, 80333 Munich, Germany}
\affiliation{Munich Center for Quantum Science and Technology (MCQST), 80799 Munich, Germany}

\author{Lode Pollet${}^{\orcidlink{0000-0002-7274-2842}}$}
\affiliation{Department of Physics and Arnold Sommerfeld Center for Theoretical Physics (ASC), Ludwig Maximilian University of Munich, 80333 Munich, Germany}
\affiliation{Munich Center for Quantum Science and Technology (MCQST), 80799 Munich, Germany}

\author{Jad C.~Halimeh${}^{\orcidlink{0000-0002-0659-7990}}$}
\email{jad.halimeh@lmu.de}
\affiliation{Department of Physics and Arnold Sommerfeld Center for Theoretical Physics (ASC), Ludwig Maximilian University of Munich, 80333 Munich, Germany}
\affiliation{Max Planck Institute of Quantum Optics, 85748 Garching, Germany}
\affiliation{Munich Center for Quantum Science and Technology (MCQST), 80799 Munich, Germany}
\affiliation{Department of Physics, College of Science, Kyung Hee University, Seoul 02447, Republic of Korea}

\date{\today}

\begin{abstract}
Mapping finite-temperature dynamical phase diagrams of quantum many-body models is a necessary step towards establishing a framework of far-from-equilibrium quantum many-body universality. However, this is quite difficult due, in part, to the severe challenges in representing the volume-law entanglement that is generated under nonequilibrium dynamics at finite temperatures. 
Here, we address these challenges with an efficient equilibrium quantum Monte Carlo (QMC) framework for computing the finite-temperature dynamical phase diagram.
Our method uses energy conservation and the self-thermalizing properties of ergodic quantum systems
to determine observables at late times after a quantum quench.
We use this technique to chart the dynamical phase diagram of the $2+1$D quantum Ising model generated by quenches of the transverse field in initial thermal states.
Our approach allows us to track the evolution of dynamical phases as a function of both the initial temperature and transverse field.
Surprisingly, we identify quenches in the ordered phase that cool the system as well as an interval of initial temperatures where it is possible to quench from the paramagnetic (PM) to ferromagnetic (FM) phases. 
Our method gives access to dynamical properties without explicitly simulating unitary time evolution, and is immediately applicable to other lattice geometries and interacting many-body systems. 
Finally, we propose a quantum simulation experiment on state-of-the-art digital quantum hardware to directly probe the predicted dynamical phases and their real-time formation.
\end{abstract}

\maketitle

\begin{figure}
    \centering
    \includegraphics[width=0.95\linewidth]{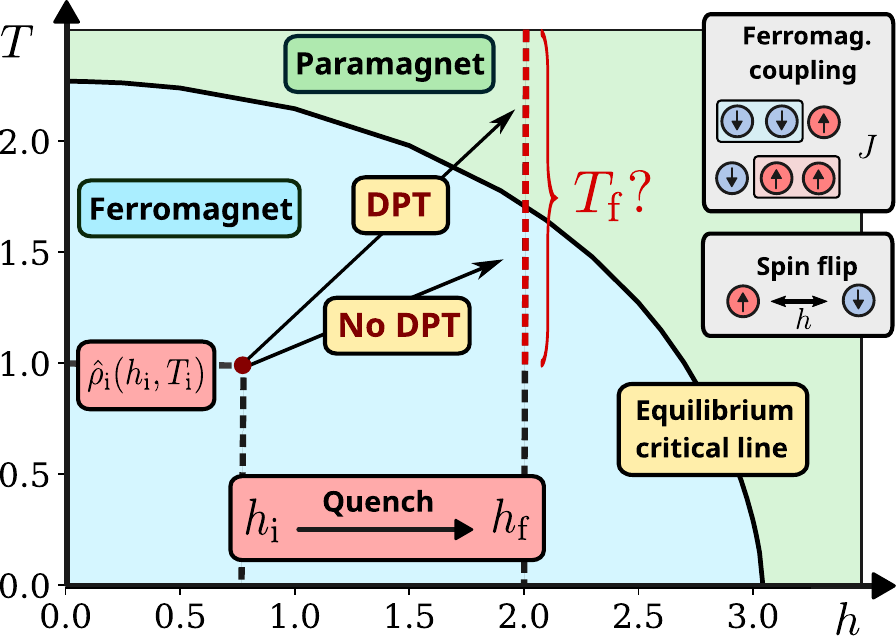}
    \caption{Illustration of our method using the $2+1$D quantum Ising model on a square lattice. The system is prepared in a thermal ensemble $\hat{\rho}(h_\text{i},T_\text{i})$ at transverse-field strength $h_\text{i}$ and temperature $T_\text{i}$, which is then quenched by instantly changing the transverse-field value to $h_\text{f}$. 
    Using QMC sampling, we determine the temperature $T_\text{f}$ of the thermal ensemble describing the long-time steady state. Knowledge of $h_\text{f}$, $T_\text{f}$ and the equilibrium phase diagram allows us to ultimately map the full finite-temperature dynamical phase diagram.
    }
    \label{fig:overview}
\end{figure}

\textbf{\textit{Introduction.---}}
Understanding universality far from equilibrium remains a central challenge in quantum many-body physics \cite{Polkovnikov2011Nonequilibriumdynamics,Eisert2015QuantumManyBodySystems}. While equilibrium phase transitions are well characterized \cite{cardy1996scaling,sachdev2011quantumphasetransitions}, the structure of long-time steady states following quantum quenches—especially at finite temperature and in two spatial dimensions—remains significantly less understood \cite{Mori2018ThermalizationAndPrethermalization,Mitra2018QuantumQuenchDynamics,Mallaya2019PrethermalizationAndThermalization}. Dynamical phase transitions (DPTs), defined through nonanalytic changes in post-quench behavior (see Fig.~\ref{fig:overview}) \cite{Sciolla2010QuantumQuenchesOffEquilibrium,Sciolla2011Dynamicaltransitionsandquantumquenches,Smacchia2015ExploringDynamicalPhaseTransitions,Zunkovic2016DynamicalPhaseTransitions,Halimeh2017Prethermalizationandpersistentorder,Halimeh2017DynamicalPhaseDiagram,Homrighausen2017Anomalousdynamicalphasetransition,Zunkovic2018DynamicalQuantumPhaseTransitions}, provide a natural framework to address this problem. However, determining finite-temperature dynamical phase diagrams of interacting $2+1$D systems is computationally demanding: exact diagonalization (ED) \cite{Sandvik2010ComputationStudiesofQuantumSpinSystems} is restricted to small lattices, and tensor network methods \cite{Schollwock2011DensitymatrixRenormalizationGroup,Orus2019TensorNetworksComplex,Paeckel2019TimeevolutionMethodsMatrixproduct,Montangero2018IntroductionTensorNetwork} face severe limitations due to the volume-law entanglement growth of thermal states evolving out of equilibrium. 
This challenge is particularly relevant in settings ranging from quantum simulators \cite{Bloch2012QuantumSimulationUltracoldQuantumGases,Georgescu2014QuantumSimulation,Gross2017QuantumSimulations,Altman2021QuantumSimulators,Alexeev2021QuantumComputerSystemsScientificDiscovery} to high-energy physics \cite{Weinberg1995QuantumTheoryFields,Gattringer2009QuantumChromodynamicsLattice,Zee2003QuantumFieldTheory}, where understanding how strongly interacting quantum systems thermalize—such as in higher-dimensional gauge theories—remains an outstanding problem across lattice and quantum computing approaches \cite{Ellis2003QCDColliderPhysics,Berges2021QCDthermalization,Bauer2023QuantumSimulationHighEnergy,DiMeglio2024QuantumComputingHighEnergy,Halimeh2025QuantumSimulationOutofequilibrium,Mueller:2024mmk,Davoudi:2024osg}.

In thermalizing systems obeying the Eigenstate Thermalization Hypothesis (ETH), long-time observables after a quench are determined by conserved quantities, such as the energy
\cite{Deutsch1991,Srednicki1994ChaosAndQuantumThermalization,Rigol2006HardcoreBosons,Rigol2008Thermalization,DAlessio2016FromQuantumChaos,Deutsch2018ETH}. 
This enables a route to mapping dynamical phase diagrams that bypasses explicit real-time evolution: for a quench from a thermal initial state of $\hat{H}_\text{i}$ to a final Hamiltonian $\hat{H}_\text{f}$, the conserved post-quench energy fixes the temperature $T_\text{f}$ of the state at late times.   
The dynamical phase is identified from the corresponding point in the equilibrium phase diagram of $\hat{H}_\text{f}$. We develop and implement this strategy to map finite-temperature dynamical phase boundaries using only static equilibrium simulations.

As a paradigmatic example, we consider the transverse-field Ising model on a square lattice. Although its finite-temperature equilibrium phase diagram is well established \cite{Hesselmann_2016, Friedman1978IsingModelWithTransverseField}, its finite-temperature nonequilibrium behavior remains largely unexplored. We obtain the full finite-temperature dynamical phase diagram following quenches of the transverse field from thermal initial states and find qualitative deviations from equilibrium behavior. Notably, finite-temperature quenches can drive the system from the PM into the FM phase, and ordered phases can be destroyed by either increasing or decreasing the transverse field. Near the quantum critical regime, dynamical ordering is strongly suppressed, highlighting the role of critical fluctuations in nonequilibrium phase formation. We benchmark our predictions against real-time unitary dynamics obtained from ED and tree tensor network (TTN) simulations \cite{tagliacozzo_simulation_2009,murg_simulating_2010}, finding good agreement on system sizes that are accessible to direct time evolution.

Because the steady-state phase structure is fully determined by conserved quantities in ergodic systems, our framework establishes a direct bridge between equilibrium thermodynamics and long-time nonequilibrium phases. This provides a scalable route to finite-temperature dynamical phase diagrams in dimensions where explicit real-time simulations are otherwise prohibitive, and opens the door to systematic studies of nonequilibrium universality in interacting quantum matter. Quantum simulators can then probe the transient unitary dynamics within and across these dynamical phases, including extracting dynamical scaling behavior near the corresponding critical lines, without the need to first reconstruct the dynamical phase diagram from scratch.

\textbf{\textit{Dynamic predictions from quantum Monte Carlo.---}}
We consider a thermal ensemble $\hat{\rho}_\text{i}=e^{-\hat{H}_\text{i}/T_\text{i}}/\Tr\big\{e^{-\hat{H}_\text{i}/T_\text{i}}\big\}$ with a given initial Hamiltonian $\hat{H}_\text{i}$ and an initial temperature $T_\text{i}$, where we have set $\hbar=k_\text{B}=c=1$ throughout. Assume a quantum quench is performed on this initial state with the Hamiltonian $\hat{H}_\text{f}$. This results in the time-evolved state $\hat{\rho}(t)=e^{-i\hat{H}_\text{f}t}\hat{\rho}_\text{i}e^{i\hat{H}_\text{f}t}$. In a thermalizing system, the long-time steady state $\hat{\rho}_\text{ss}=\lim_{\tau\to\infty}\frac{1}{\tau}\int_0^\tau dt\, \hat{\rho}(t)$ resulting from this quench is expected to be locally equivalent to the thermal ensemble $\hat{\rho}_\text{f}=e^{-\hat{H}_\text{f}/T_\text{f} }/\Tr\big\{e^{-\hat{H}_\text{f}/T_\text{f} }\big\}$ at temperature $T_\text{f}$.
Our method is illustrated in Fig.~\ref{fig:overview}.

Energy conservation fixes the post-quench energy:
\begin{align}\label{eq:cond}
E_\text{q}=\Tr\big\{\hat{\rho}_\text{i}\hat{H}_\text{f}\big\}=\Tr\big\{\hat{\rho}_\text{f}\hat{H}_\text{f}\big\}.
\end{align} 
This is an implicit equation with $T_\text{f}$ as the only unknown variable, which can then be solved using, e.g., the Newton--Raphson method or a bisection search. 
For small system sizes, it suffices to use ED to solve Eq.~\eqref{eq:cond}, while for relevant larger system sizes, we turn to QMC sampling; see Supplemental Material (SM) for details \cite{SM}. Upon obtaining $T_\text{f}$, and based on the value of $h_\text{f}$, we can look to the equilibrium phase diagram \cite{Hesselmann_2016} of the TFIM to determine the phase of the system after thermalization.

\textbf{\textit{$2+1$D quantum Ising model.---}}
Although our method is general and can be applied to generic interacting quantum many-body models in any spatial dimension, we will demonstrate it in this work by focusing on the transverse-field Ising model (TFIM) on an $L\times L$ square lattice with periodic boundary conditions. This paradigmatic model is described by the Hamiltonian
\begin{align}\label{eq:TFIM}
    \hat{H} = -J \sum_{\langle \mathbf{i},\mathbf{j}\rangle} \hat{\sigma}^z_\mathbf{i}\hat{\sigma}^z_\mathbf{j} - h\sum_\mathbf{i} \hat{\sigma}^x_\mathbf{i}, 
\end{align}
where $J=1$ is the FM coupling, which we set to unity throughout, $h$ is the transverse-field strength, $\hat{\sigma}^{x,z}_\mathbf{i}$ are Pauli matrices defined on site $\mathbf{i}$, and $\langle \mathbf{i},\mathbf{j}\rangle$ denotes nearest neighbors counted only once. As per our notation, the initial (final) Hamiltonian $\hat{H}_\text{i}$ ($\hat{H}_\text{f}$) corresponds to the transverse-field strength in Eq.~\eqref{eq:TFIM} being equal to its initial (final) value $h_\text{i}$ ($h_\text{f}$).
The equilibrium phase diagram and critical exponents of the $2+1$D TFIM are well known \cite{sachdev2011quantumphasetransitions,Chang:2024whx}. It has an equilibrium quantum critical point $h_\text{c}^\text{e}\approx3.044$ at zero temperature \cite{deJongh1998criticalbehavior,Bloete2002clustermontecarlo} and a critical temperature $T_\text{c}^\text{e}=\frac{2}{\ln (1+\sqrt{2})}$ at zero transverse-field strength \cite{Onsager1944crystalstatistics}; see Fig.~\ref{fig:overview}. However, its finite-temperature dynamical phase diagram had not been mapped until this work, despite progress at $T_\text{i}=0$ \cite{Hashizume2022DynamicalModel}. The $2+1$D TFIM is nonintegrable and is expected to thermalize in the long-time limit \cite{Bla2016,Mondaini_2016,Mondaini2017Eigenstatethermalization,Chiba_2025}.

\begin{figure}
    \centering
    \includegraphics[width=\linewidth]{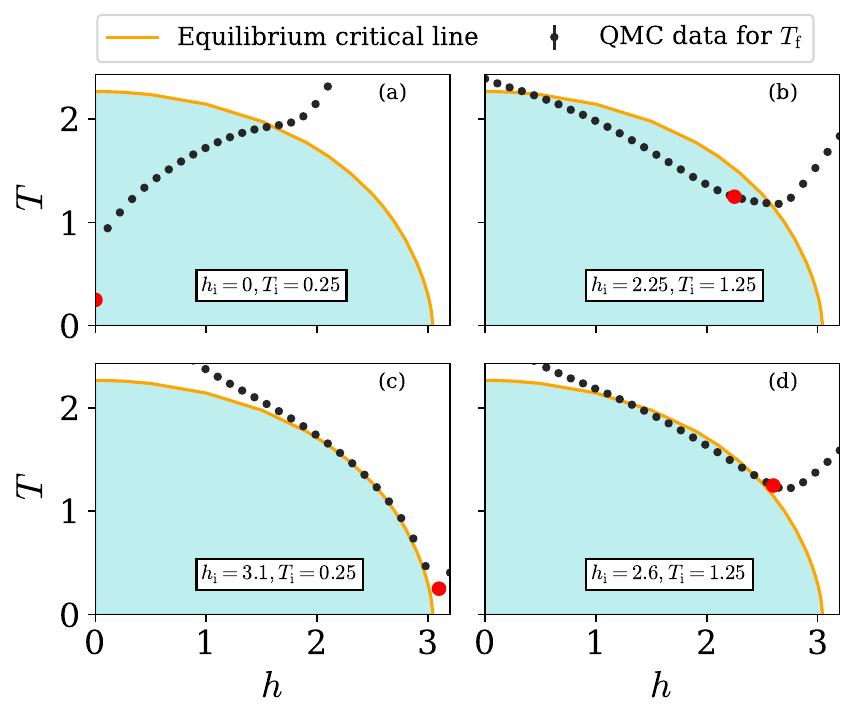}
    \caption{The $2+1$D quantum Ising model on a square lattice is prepared in the thermal ensemble $\hat{\rho}_\text{i}(h_\text{i},T_\text{i})$ (red dot). The system is subsequently quenched by instantly changing the transverse-field value to $h_\text{f}$. The temperature $T_\text{f}$ (black circles) of the long-time steady state of the quenched system is obtained from the conservation of the quench energy. 
    The data is obtained by solving the implicit Eq.~\eqref{eq:cond} for $24\times24$ systems using the bisection method and QMC simulations; see~\cite{SM}. 
    The equilibrium critical line (orange line) 
    is from Ref.~\cite{Hesselmann_2016}, where the cyan region corresponds to the FM phase.}
    \label{fig:four_Tf}
\end{figure}

\textbf{\textit{QMC results.---}} 
We find that for most cases considered in our work, a lattice of size $24\times24$ approaches the thermodynamic limit well \cite{SM}. Furthermore, this system size facilitates reasonable runtimes, thereby allowing fine scans in $T_\mathrm{i}$ and $h_\mathrm{f}$.
Figure~\ref{fig:four_Tf} shows our computed values of $T_\text{f}$ for quenched systems with different initial conditions $(h_\text{i},T_\text{i})$. The intersection of the interpolated line between the $T_\text{f}$ data points with the equilibrium phase diagram gives the dynamical critical value of the transverse field at which the DPT occurs. Starting deep in the FM phase (small $h_\text{i}$ and $T_\text{i}$), relatively large quenches are required to drive the system to the PM phase, as demonstrated in Fig.~\ref{fig:four_Tf}(a). Starting within the FM phase but at larger values of $h_\text{i}$ and $T_\text{i}$, we find a more interesting behavior resulting in two dynamical critical points, as displayed in Fig.~\ref{fig:four_Tf}(b). The first occurs at a slightly higher value of the transverse field $h_\text{c}^{\text{d},1}\gtrsim h_\text{i}$ due to the proximity of the initial state to the PM phase. The second is at a much lower value $h_\text{c}^{\text{d},2}\ll h_\text{i}$. Such a large quench deep into the FM phase injects enough energy into the system to drive it into the PM phase.
Interestingly, we observe \textit{cooling quenches}: Quenching to $h_\text{f}\in (2.25,2.75)$ leads to a final temperature $T_\text{f}<T_\text{i}$.
This does not violate energy conservation, as temperature is defined relative to the spectrum of the post-quench Hamiltonian. The same conserved energy can correspond to different final temperatures depending on the density of states of $\hat{H}_\text{f}$, highlighting how quench dynamics can redistribute energy in a way that effectively cools the system.

\begin{figure}
    \centering
    \includegraphics[width=\linewidth]{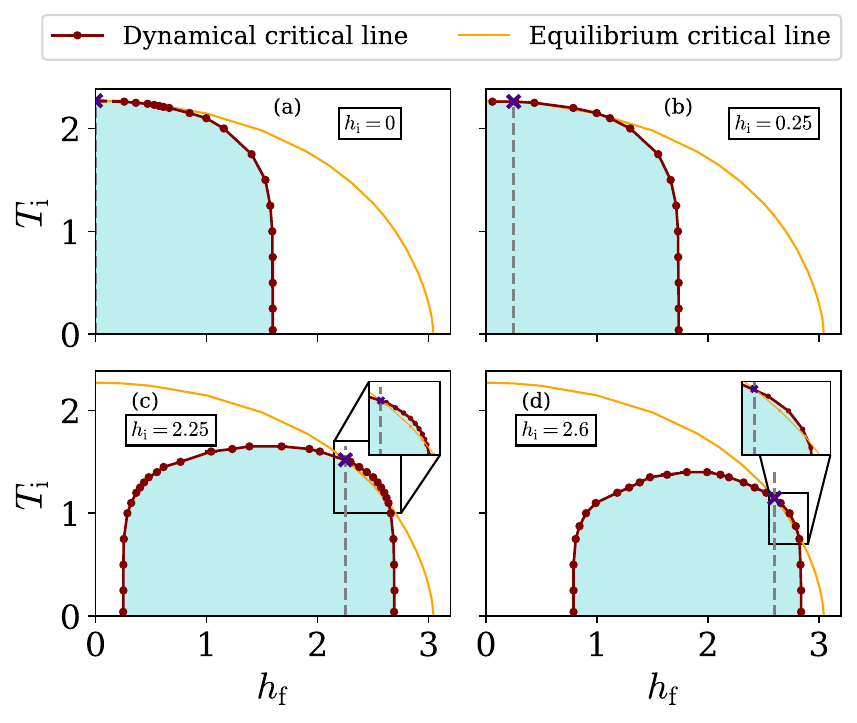}
    \caption{Finite-temperature dynamical phase diagrams (red line denotes $h_\text{c}^\text{d}(h_\text{i},T_\text{i}$); FM dynamical phase in cyan) for different values of $h_\text{i}$ (grey dashed line). The cyan region corresponds to systems with initial temperature $T_\mathrm{i}$, whose long-time steady states after a quench with a Hamiltonian with $h_\mathrm{f}$ will be in the FM phase. The orange line corresponds to the equilibrium critical line for systems with transverse field $h_\text{f}$ and temperature $T_\text{i}$. The indigo cross marks the equilibrium critical temperature at $h_\text{i}$. The lowest considered temperature is $T_\mathrm{i} = \frac{1}{L}$. The error bars on the dynamical critical line are obtained from the different intersection points of the upper and lower bound of $T_\mathrm{f}$ with the equilibrium critical line.}
    \label{fig:four_dynam_phase}
\end{figure}

Another intriguing behavior occurs when initializing the system in the PM phase close to the $T_\text{i}=0$ quantum critical point, as shown in Fig.~\ref{fig:four_Tf}(c). 
Despite being initialized so close to the FM phase, the system never dynamically thermalizes into the FM phase. 
The reason is that the mass gap $\Delta$ of the model is smaller than the initial temperature ($T_\text{i}\gtrsim\Delta)$,
which leads to thermal fluctuations dominating the gap scale \cite{sachdev2011quantumphasetransitions}, keeping the system in the PM phase after any quench.
Staying close to the equilibrium critical line but away from the quantum critical regime, the picture fundamentally changes, as shown in Fig.~\ref{fig:four_Tf}(d). Here, two dynamical critical points emerge. The first is slightly smaller, $h_\text{c}^{\text{d},1}\lesssim h_\text{i}$, and the system dynamically thermalizes to the FM phase, exemplifying a process of dynamically induced order after starting in a PM state. Finite-size scaling analysis indicates that this persists in the thermodynamic limit \cite{SM}. Quenching deeper into the FM phase injects more energy into the system, leading to a second much smaller dynamical critical point, $h_\text{c}^{\text{d},2}< h_\text{i}$, where the system is dynamically driven into the PM phase. As in the case of Fig.~\ref{fig:four_Tf}(b), we also find cooling quenches for $h_\text{f}\in(2.6,2.75)$.

Fixing the value of $h_\text{i}$ and performing these calculations for different initial temperatures $T_\text{i}$, we can determine the corresponding dynamical critical points as well as the corresponding finite-temperature dynamical phase diagram. Figure~\ref{fig:four_dynam_phase} shows four such diagrams for different $h_\text{i}$. In all cases, the critical line has a weak dependence on the initial temperature for low values of $T_\text{i}$. The smallest temperature simulated in QMC is $T_\mathrm{i} = \frac{1}{24}$. For zero and small $h_\text{i}$, shown in Fig.~\ref{fig:four_dynam_phase}(a,b), the dynamical critical line resembles a rescaled version of the equilibrium one with a transition at a lower value of the transverse field due to the heating from the quench. For example, at $h_\text{i}=0$ and $T_\mathrm{i} = \frac{1}{24}$, we find that $h_\mathrm{c}^\mathrm{d} \approx 1.6$. For higher values of $h_\text{i}$, we find that the dynamical phase diagram qualitatively differs from its equilibrium counterpart, and the FM phase can be left by increasing or decreasing the value of $h_\text{f}$.

An interesting feature in the dynamical phase diagrams of Fig.~\ref{fig:four_dynam_phase}(c,d) is the existence of a region where the dynamical critical line is \textit{above} its equilibrium counterpart, e.g., the region within $T_\text{i}\in(1.25,1.5)$ and $h_\text{f}\in(2.3,2.6)$ in Fig.~\ref{fig:four_dynam_phase}(c). This is a direct consequence of the cooling quenches where the thermal ensemble at $(T_\text{f},h_\text{f})$ is colder than that at $(T_\text{i},h_\text{f})$ since $T_\text{f}<T_\text{i}$, giving rise to situations where the latter (former) ensemble is in the PM (FM) phase.

It is important to compare Fig.~\ref{fig:four_dynam_phase}(a) with previous work, where we find a tighter bound on $h_\text{c}(h_\text{i}=0,T_\text{i}=0)$ relative to Refs.~\cite{Hashizume2022DynamicalModel,borla2026microscopicdynamicsfalsevacuum}, which studied a system size of $\infty\times6$ in matrix product states and $16\times16$ in TTNs, respectively. In these works, an upper bound on $h_\text{c}(h_\text{i}=0,T_\text{i}=0)$ was determined by searching for the smallest value of $h_\text{f}$ at which the order parameter dynamically crosses zero within the accessible evolution times, considering this a certified signature of an infinite-time PM steady state. However, this certainly does not exclude that smaller values of $h_\text{f}$ can still lead to an infinite-time PM steady state. Indeed, in the case of the $1+1$D TFIM with short-range interactions, any quench $h_\text{i}=0\to h_\text{f}<h_\text{c}^{\text{e},1+1\text{D}}$ at $T_\text{i}=0$ will lead to an infinite-time PM phase with the order parameter asymptotically decaying to, but never crossing, zero \cite{Calabrese2011quantumquenchinthetransverse-fieldisingchain,Calabrese2012quantumquenchinthetransverse-fieldisingchainI,Calabrese2012quantumquenchinthetransverse-fieldisingchainII,Halimeh2020QuasiparticleOrigin}.

\begin{figure}
    \centering
    \includegraphics[width=0.95\linewidth]{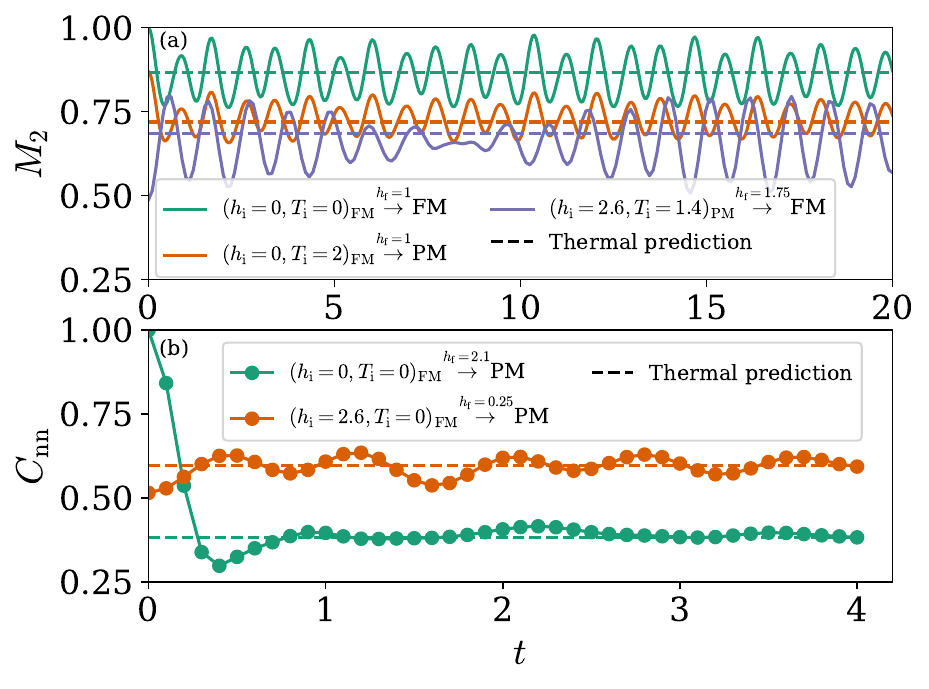}
    \caption{Real-time evolution of observables after the quantum quench: (a) second moment of magnetization on a $4\times4$ lattice from ED, (b) nearest-neighbor correlator on a $8\times8$ lattice from TTNs. In TTNs, we restrict ourselves to $T_\text{i} = 0$, in ED, additionally, finite $T_\text{i}$ is considered. The real-time dynamics is compared with the predicted properties of the long-time steady state of the quenched systems (dashed lines). The temperature of the long-time steady state is obtained from solving Eq.~\eqref{eq:cond} implicitly using QMC on the same-sized lattices.  For the $T_\mathrm{i} = 0$ predictions, the quench energy is obtained from TTNs, after which the final temperature is fixed using QMC.}
    \label{fig:dynamics}
\end{figure}

\textbf{\textit{Real-time dynamics.---}}
In order to benchmark our QMC-based method, we now compute the unitary dynamics of the considered quenches using ED for finite-temperature ensembles. For $T_\text{i}=0$, we use both ED and TTNs, with the latter employing the time-dependent variational principle (TDVP) \cite{Haegeman2011TimeDependentVariationalPrinciple,Haegeman2016UnifyingTimeEvolution,Vanderstraeten2019TangentSpaceMethods}. In the case of $h_\text{i} = 0$ and $T_\text{i} = 0$, we start in the fully up-polarized ground state. In ED, we calculate the second moment of magnetization $M_2=\frac{1}{L^4}\Tr\big\{\hat{\rho}(t)\sum_{\mathbf{i},\mathbf{j}} \hat{\sigma}_{\mathbf{i}}^z\hat{\sigma}_{\mathbf{j}}^z\big\}$, which is a measure of long-range order. In TTNs, obtaining long-range correlations is computationally costly, so instead we calculate the nearest-neighbor correlator $C_\text{nn}= \frac{1}{L^2}\sum_{\langle \mathbf{i},\mathbf{j}\rangle}\Tr\big\{\hat{\rho}(t)\hat{\sigma}_\mathbf{i}^z \hat{\sigma}_\mathbf{j}^z\big\}$.  
The results are shown in Fig.~\ref{fig:dynamics}.
We find good agreement between thermal predictions and unitary dynamics for the $8 \times 8$ systems studied using TTNs. For small quenches in the $4\times4$ system, the agreement between the thermal prediction and the time-evolved values is good for both zero and finite $T_\mathrm{i}$. For quenches with $h_\text{f}$ close to the dynamical critical value of the transverse field, small discrepancies arise. In these cases, the systems after the quench are near criticality and critical slowing down is expected. Therefore, the discrepancies could arise due to thermalization timescales being larger than the simulable times. Furthermore, finite-size effects are expected to be larger near criticality.

\textbf{\textit{Proposal for quantum simulation.---}} 
QMC gives access to the late-time equilibrium properties following a quench but does not give insight into the mechanisms that drive thermalization. 
This requires tracking how observables evolve in real-time.
Such information not only checks the validity of the ETH, as was done in the previous section, but can also probe dynamical universality~\cite{RevModPhys.49.435}.
For quenches to the critical line $h_\text{i}\to h_\text{c}^\text{d}$, dynamical universality predicts that 
observables relax to thermal equilibrium in a manner that is independent of $(h_\text{i},T_\text{i})$, and with universal scaling governed by the static and dynamical critical exponents of the phase transition.
The zero (finite) temperature transition belongs to the $3$D ($2$D) classical Ising universality class and energy-conserving quenches are classified by Model C of Ref.~\cite{RevModPhys.49.435}.
Access to real-time data would directly test the assumptions underlying dynamical universality, see, e.g., Refs.~\cite{Samajdar:2024fjc,Balducci:2025ztg,Khasseh:2020yke,Manovitz:2024hif}.

Computing real-time correlation functions is notoriously difficult at scale.
Tensor networks cannot represent the highly entangled quantum states that are produced in finite-temperature quenches~\cite{Calabrese:2005in,Schuch:2008zza,Haghshenas:2025euj} while ED is limited to small lattices.
Recently, digital quantum computers have emerged as a promising alternative for simulating quantum many-body dynamics~\cite{Altman2021QuantumSimulators,Alexeev2021QuantumComputerSystemsScientificDiscovery}.
Simulations using quantum computers face a different set of challenges than simulations using classical computers.
Current quantum computers are noisy, which limits computations to modest lattice sizes and early times.
This will eventually be overcome by quantum error correction but, even on a fault-tolerant quantum computer, preparing an initial thermal ensemble is nontrivial.

We propose using dissipative methods, which have recently been shown to efficiently converge to the Gibbs ensemble under physically realistic assumptions~\cite{Guo:2024ycr,Chen:2025fax,Zhan:2025gof,Ding:2025ulc,Lloyd:2025cvp}. 
In their simplest form, these methods introduce additional ``bath" qubits  that have a classical Hamiltonian whose Gibbs state can easily be sampled from.
Interactions between the ancilla qubits and the TFIM spins thermalize the system to the temperature set by the bath.
Operationally, this requires time evolving under the enlarged TFIM plus bath Hamiltonian while periodically refreshing the ancilla qubits to dump energy and entropy.
The qubit refresh can be achieved with mid-circuit measurements, a feature that is supported by many current quantum computing platforms~\cite{Bluvstein:2023zmt,GoogleQuantumAIandCollaborators:2024efv,Javadi-Abhari:2024kbf,Ransford:2025ksn}.
Indeed, similar dissipative methods have recently been implemented on superconducting qubit devices to prepare TFIM ground states~\cite{Mi:2023evq,Song:2025pwd}, see also Refs.~\cite{Granet:2025sks,Li:2025rik}.
Thus, the outlook for a first demonstration of thermal quench dynamics on a near-term quantum computer is promising. 
However, quantitative predictions of how observables approach thermal equilibrium will likely require error correction.

\textbf{\textit{Summary and outlook.---}}
We have developed an efficient equilibrium quantum Monte Carlo framework for extracting the finite-temperature dynamical phase diagram of generic interacting quantum many-body models using only equilibrium thermodynamic data. As a concrete example, we have calculated the finite-temperature dynamical phase diagram of the $2+1$D transverse-field Ising model. Intriguingly, we have found initial conditions that lead to dynamical phase diagrams qualitatively different from their equilibrium counterparts, such as cooling quenches where the dynamical ordered phase extends to larger transverse fields than its equilibrium counterpart.
We have shown how starting close to the equilibrium quantum critical point in the disordered phase prevents dynamically reaching the ordered phase. 
We have benchmarked our results obtained from our QMC-based method against real-time unitary dynamics calculated using exact diagonalization and tree tensor networks, finding good agreement for the accessible system sizes.

Our method can map out finite-temperature dynamical phase diagrams of interacting quantum many-body models that are of interest to quantum simulators.  By delineating the dynamical critical boundaries, it provides a clear starting point for experimental and quantum-simulation studies of nonequilibrium scaling and transport within and across these phases.

Finally, from a numerical and phenomenological standpoint, our method can be readily applied to study the finite-temperature dynamical phase diagrams of high-energy physics models such as lattice gauge theories whose dynamic properties have been the subject of much investigation in recent quantum simulation experiments~\cite{Martinez2016RealtimeDynamicsLattice, Klco2018QuantumclassicalComputationSchwinger,Gorg2019RealizationDensitydependentPeierls, Schweizer2019FloquetApproachZ2, Mil2020ScalableRealizationLocal, Yang2020ObservationGaugeInvariance, Wang2022ObservationEmergent$mathbbZ_2$, Su2023ObservationManybodyScarring, Zhou2022ThermalizationDynamicsGauge, Wang2023InterrelatedThermalizationQuantum, Zhang2025ObservationMicroscopicConfinement, Zhu2024ProbingFalseVacuum, Ciavarella2021TrailheadQuantumSimulation, Ciavarella2022PreparationSU3Lattice, Ciavarella2023QuantumSimulationLattice-1, Ciavarella2024QuantumSimulationSU3,
Chernyshev:2025lil,
Froland:2025bqf,
Gustafson2024PrimitiveQuantumGates, Gustafson2024PrimitiveQuantumGates-1, Lamm2024BlockEncodingsDiscrete, Farrell2023PreparationsQuantumSimulations-1, Farrell2023PreparationsQuantumSimulations, 
Farrell2024ScalableCircuitsPreparing,
Farrell2024QuantumSimulationsHadron, Li2024SequencyHierarchyTruncation, 
Li:2025sgo,
Zemlevskiy2025ScalableQuantumSimulations, Lewis2019QubitModelU1, Atas2021SU2HadronsQuantum, ARahman:2022tkr, Atas2023SimulatingOnedimensionalQuantum, Mendicelli2023RealTimeEvolution, Kavaki2024SquarePlaquettesTriamond, Than2024PhaseDiagramQuantum, Angelides:2023noe,   
Mildenberger2025Confinement$$mathbbZ_2$$Lattice, Schuhmacher2025ObservationHadronScattering, Davoudi2025QuantumComputationHadron, Saner2025RealTimeObservationAharonovBohm, Xiang2025RealtimeScatteringFreezeout, Wang2025ObservationInelasticMeson,li2025frameworkquantumsimulationsenergyloss,mark2025observationballisticplasmamemory,froland2025simulatingfullygaugefixedsu2,Hudomal2025ErgodicityBreakingMeetsCriticality,hayata2026onsetthermalizationqdeformedsu2,De2024ObservationStringbreakingDynamics, Liu2024StringBreakingMechanism, Alexandrou:2025vaj}, including in two spatial dimensions \cite{Gyawali2025ObservationDisorderfreeLocalization,Cochran2025VisualizingDynamicsCharges, Gonzalez-Cuadra2025ObservationStringBreaking, Crippa2024AnalysisConfinementString,Cobos2025RealTimeDynamics2+1D}.

\bigskip
\footnotesize
\textbf{\textit{Acknowledgments.---}} The authors acknowledge formative discussions with Henrik Dreyer, Andrew Potter, and Enrico Rinaldi, and are grateful to Stefan Wessel for providing the raw data for the equilibrium critical line of the square-lattice TFIM. L.K., U.B., and J.C.H.~acknowledge funding by the Max Planck Society and the European Research Council (ERC) under the European Union’s Horizon Europe research and innovation program (Grant Agreement No.~101165667)—ERC Starting Grant QuSiGauge. Views and opinions expressed are, however, those of the author(s) only and do not necessarily reflect those of the European Union or the European Research Council Executive Agency. Neither the European Union nor the granting authority can be held responsible for them. L.K., U.B., L.P.,~and J.C.H.~acknowledge support from the Deutsche Forschungsgemeinschaft (DFG, German Research Foundation) under Germany’s Excellence Strategy – EXC-2111 – 390814868. The project/research is part of the Munich Quantum Valley, which is supported by the Bavarian state government with funds from the Hightech Agenda Bavaria. This work is part of the Quantum Computing for High-Energy Physics (QC4HEP) working group.
R.F.~acknowledges support from
the U.S.~Department of Energy QuantISED program
through the theory consortium “Intersections of QIS
and Theoretical Particle Physics” at Fermilab, from the
U.S.~Department of Energy, Office of Science, Accelerated Research in Quantum Computing, Quantum Utility
through Advanced Computational Quantum Algorithms
(QUACQ), and from the Institute for Quantum Information and Matter, an NSF Physics Frontiers Center (PHY2317110). 
R.F.~additionally acknowledges support from
a Burke Institute prize fellowship. 
\normalsize

\bibliography{biblio}

\clearpage
\pagebreak
\newpage
\onecolumngrid

\setcounter{equation}{0}
\setcounter{figure}{0}
\setcounter{table}{0}
\setcounter{page}{1}

\renewcommand{\theequation}{S\arabic{equation}}
\renewcommand{\thefigure}{S\arabic{figure}}
\renewcommand{\thetable}{S\arabic{table}}

\begin{center}
\textbf{\large Supplemental Material for ``Finite-Temperature Dynamical Phase Diagram of the $2+1$D Quantum Ising Model''} 
\end{center}

\twocolumngrid

\section{Numerical methods}
For thermal expectation values of large systems, the loop quantum Monte Carlo algorithm of the ALPS library was used~\cite{Albuquerque2007,Bauer2011,Todo2001,Troyer1998}. 
We obtain the quench energy by calculating thermal expectation values of the initial system, 
\begin{align}
    E_\mathrm{q}&=\Tr(\hat{\rho}_\text{i}\hat{H}_\text{f}) \notag
    \\ &= -J \Tr(\hat{\rho}_\text{i}\sum_{\langle \mathbf{i},\mathbf{j}\rangle} \hat{\sigma}^z_\mathbf{i}\hat{\sigma}^z_\mathbf{j}) - h_\text{f} \Tr(\hat{\rho}_\text{i}  \sum_\mathbf{i} \hat{\sigma}^x_\mathbf{i}).
\end{align}
Then, $T_\text{f}$ is determined using the bisection method. Specifically, we start by choosing an interval $(T_1,T_2)$ and calculate $E_{1,2} = \Tr(e^{-\hat{H}_\text{f}/T_{1,2}} \hat{H}_\text{f})$. Afterward, we bisect the interval $T_\text{mid} = \frac{T_1+T_2}{2}$ and calculate $E_\text{mid}$. If $(E_1-E_\text{q})(E_\text{mid}-E_\text{q}) < 0$, then $T_\text{f} \in (T_1,T_\text{mid})$ else $T_\text{f} \in (T_\text{mid},T_2)$. We choose the respective interval as new initial interval and iterate until, $|E_\text{mid}-E_\text{q}| \le \Delta E_\text{mid} + \Delta E_\text{q}$, where $\Delta E$ are the statistical errors due to a finite number of samples in the QMC simulations. Due to these errors, $T_\text{f}$ can not be exactly determined. To get an estimate of the error in $T_\text{f}$, after the bisection, additional simulations are run with temperatures around $T_\text{f}$. For the bisection and the error estimation, several QMC simulations per quench need to be performed, making runtime the limiting factor.

Tree Tensor Networks were implemented in the QuantumTeaLeaves library~\cite{qtealeaves,qredtea}. TTNs are inherently $1$d, therefore, the $2+1$D system is mapped to a spin chain, which comes at the cost of nearest-neighbor interactions becoming long-range. To still properly capture the interactions, the bond dimension needs to grow with the system size and entanglement. This increases the needed memory and runtimes, limiting the simulable system sizes, especially for large quenches. For the time evolution in TTNs, a one-tensor TDVP update is used. The TTN simulations were performed on a NVIDIA A40 GPU with $46$ GB of memory. 

For ED, the QuSpin library was used \cite{Weinberg_2017}. Using the translation and parity symmetry, we can decompose the TFIM Hamiltonian in different symmetry sectors, rendering it block diagonal, 
\begin{align}
    \hat{H} = \bigoplus_q \hat{H}_q. 
\end{align}
The Gibbs ensemble as well as the time evolution operator are also block-diagonal allowing us to rewrite, 
\begin{align}\label{eq:ED}
    \Tr(e^{i\hat{H}_\mathrm{f}t} \hat{\rho}_\mathrm{i} e^{-i\hat{H}_\mathrm{f}t}\hat{O})=  \frac{\sum_q\Tr(e^{-i\hat{H}_{\mathrm{f},q}t} e^{-\hat{H}_{\mathrm{i},q}/T_\text{i}} e^{i\hat{H}_{\mathrm{f},q}t}\hat{O}_q)}{\sum_q\Tr(e^{-\hat{H}_{\mathrm{i},q}/T_\text{i}})},
\end{align}
for all observables with the same block-diagonal structure. Each symmetry block is then exactly diagonalized. For a $4\times 4$ lattice, the largest symmetry sector has dimension $4080\times4080$, using double point precision the eigenvectors need $\approx 133$ MB of memory. The Hilbert space grows exponentially, and for $5\times 5$ lattices the largest sector already has dimension $1342176 \times 1342176$ requiring $\mathcal{O}(\mathrm{TB)}$ of memory, making simulations unfeasible. 

\begin{figure}
    \centering
    \includegraphics[width=0.95\linewidth]{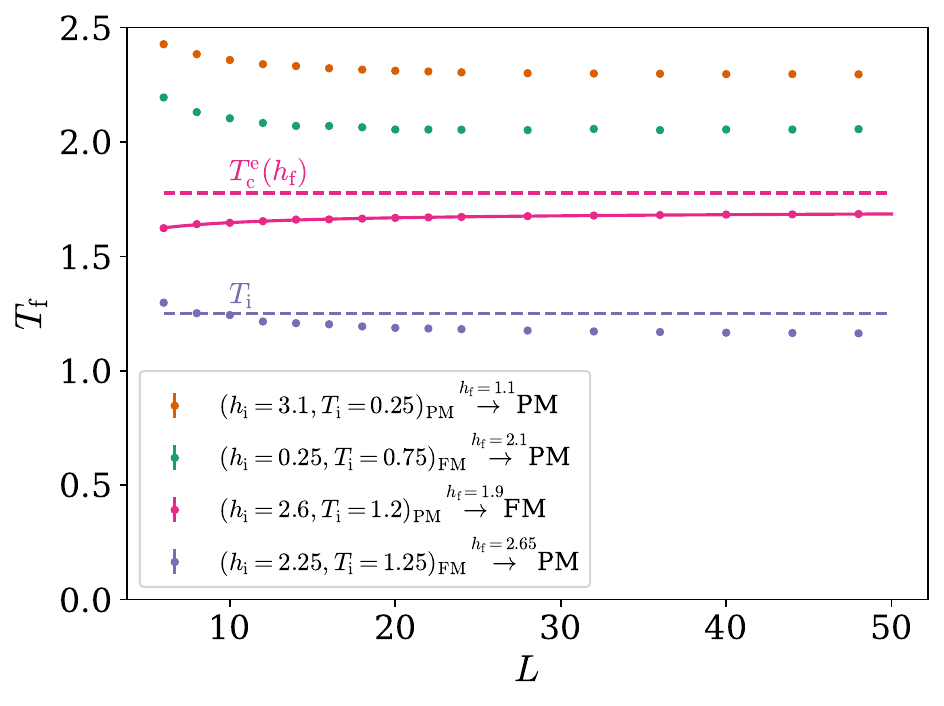}
    \caption{Finite-size scaling of the temperatures of the long-time steady states for four different quantum quenches. The color of each dataset corresponds to a different quench (see legend). The two upper quenches (green, orange) were selected because of their pronounced finite-size effects. The third quench from the top (pink), takes a PM initial state to an FM long-time steady state; we indicate the equilibrium critical temperature corresponding to $h_\text{f}$  (pink dashed line). Fitting a power law with constant offset to the data has small residuals. According to this fit, PM to FM transitions persist even in the thermodynamic limit. The dataset at the bottom (purple) corresponds to a cooling quench. We indicate the initial temperature of this quench (purple dashed line), visualizing the cooling effect becoming more pronounced with larger system size.}
    \label{fig:FiniteSizeTf}
\end{figure}
\begin{figure}
    \centering
    \includegraphics[width=0.9\linewidth]{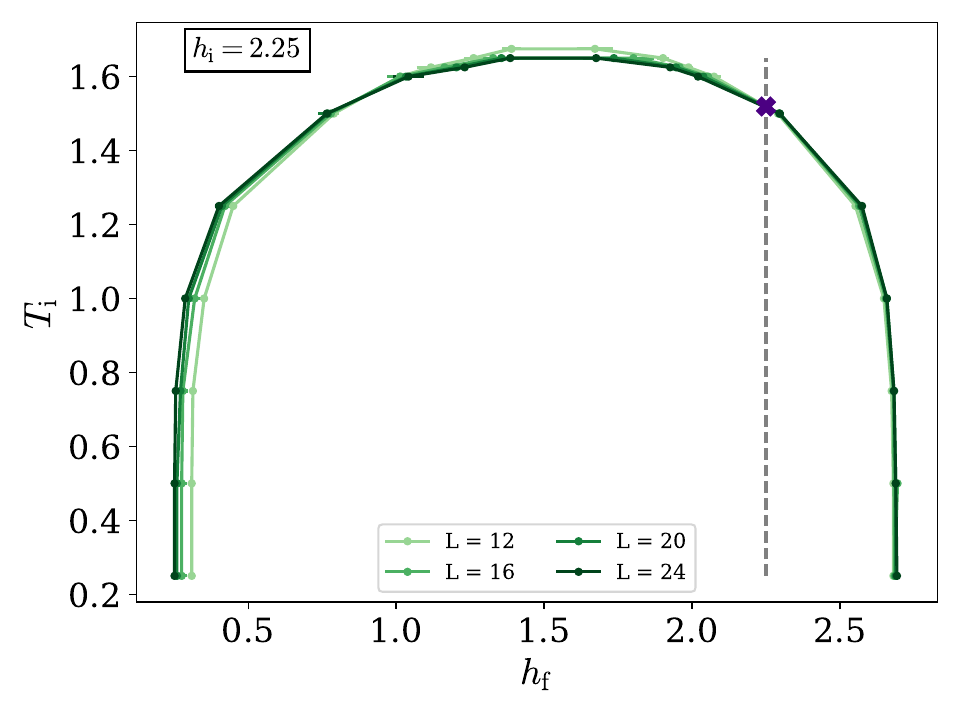}
    \caption{Dynamical phase diagram for $h_\text{i} = 2.25$ obtained from different lattice sizes. Good convergence is already reached for systems larger than $20 \times 20$. Consistent with Fig.~\ref{fig:FiniteSizeTf}, the region in which PM to FM quenches are possible gets smaller for larger systems, but persists.}
    \label{fig:FiniteSizeCriticalLine}
\end{figure}  
\begin{figure}
    \centering
    \includegraphics[width=0.95\linewidth]{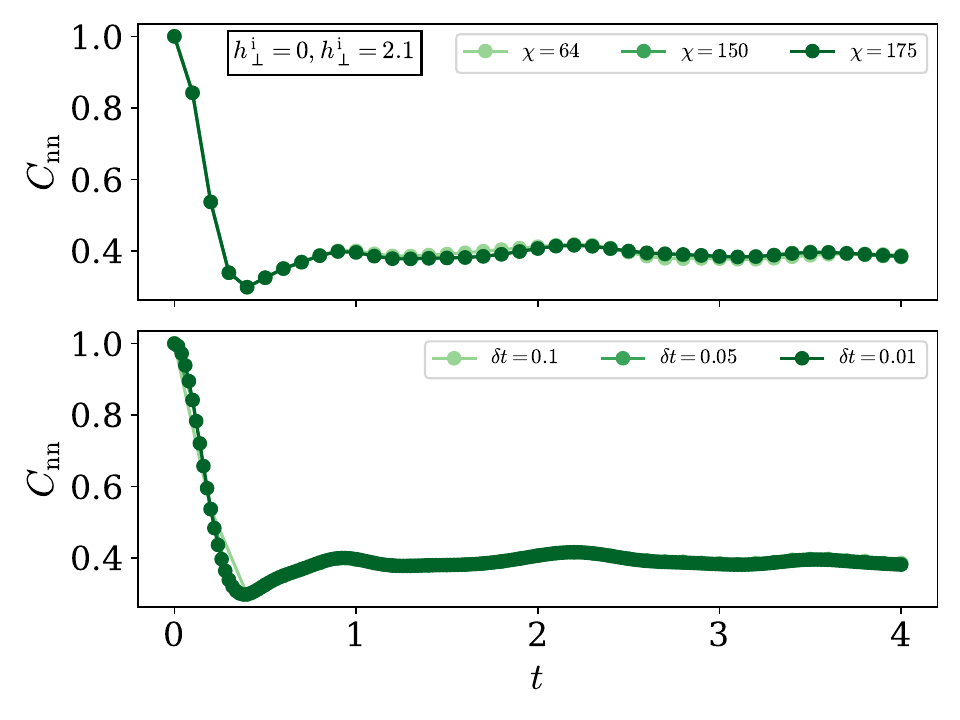}
    \caption{Convergence analysis of the TTN real-time dynamics of the nearest-neighbor correlator. In the upper plot, the time-step is kept constant at $\delta t = 0.05$, while in the lower plot the bond dimension is kept constant at $\chi = 150$. We observe good convergence for $\chi = 150$ and $\delta t = 0.05$.}
    \label{fig:TTN_convergence}
\end{figure}

\section{Finite-size scaling of the final temperature}
Both the energy density and the transverse-field magnetization scale with system size. The nonleading-order corrections to the finite-size scaling differ for different values of $h$. This complicates the finite-size scaling of $T_\mathrm{f}$, especially for large quenches. Still, we see that for large enough system sizes, $T_\text{f}$ converges. Figure~\ref{fig:FiniteSizeTf} shows the finite-size scaling of the final temperature for two quenches with large differences between $h_\mathrm{i}$ and $h_\mathrm{f}$ (orange and green datasets). As well as the finite-size effects on cooling quenches(purple dataset) and transitions from the PM to the FM phase (pink dataset). The decrease in temperature in cooling quenches becomes more pronounced for larger systems. The temperature range in which quenches from the PM to the FM phase are allowed becomes smaller for larger systems, but these transitions are observed for all considered lattice sizes (up to $60\times60$). Assuming $T_\text{f}$ converges to a finite value, we use a least-squares method to fit the data to the function $T_\mathrm{f}(L) = aL^{-b}+c$, where $a,b,c$ are the to-be-determined fit parameters. The fit is in good agreement with the data, showing a root mean square error of $\mathrm{RMSE} = 0.00097$. We find that $c = 1.70 < T_\mathrm{c}^\mathrm{e}$, suggesting that even in the thermodynamic limit, the PM to FM quenches persist.  

The dynamical critical values of $h_\text{f}$ depend on $T_\mathrm{f}$ and hence on system size. The resulting differences in the dynamical phase diagram for different system sizes become small for systems larger than $20\times20$. 

Fig.~\ref{fig:FiniteSizeCriticalLine} shows the finite-size effects on the dynamical phase diagram. Showing that for large quenches, finite-size effects are more pronounced, making the region with small $h_\text{f}$ more dependent on system size. Furthermore, the temperature regime in which quenches from the PM to the FM phase are allowed shrinks with larger system sizes.   

\section{Convergence analysis for Tree Tensor Networks} 
For the TTN simulations, good convergence is reached at bond dimensions $\chi = 150$ and time-step size of $\delta t = 0.05$. Figure~\ref{fig:TTN_convergence} shows the nearest-neighbor correlation, for different bond dimensions and time steps.

\end{document}